\documentclass[10pt,journal]{IEEEtran}
\usepackage{amsmath,amssymb,amsfonts,pifont}
\usepackage{algorithmic}
\usepackage{graphicx}
\usepackage{booktabs}
\usepackage{textcomp}
\usepackage{color}
\usepackage{multirow}
\usepackage{multicol}
\usepackage{url}
\usepackage[colorlinks,linkcolor=blue]{hyperref}
\usepackage{array}
\usepackage{xcolor}
\usepackage{bm}
\usepackage{float}
\usepackage[ruled, lined, linesnumbered, commentsnumbered, longend]{algorithm2e}
\usepackage[caption=false, font=footnotesize]{subfig}
\usepackage[numbers]{natbib}
\def\BibTeX{{\rm B\kern-.05em{\sc i\kern-.025em b}\kern-.08em
    T\kern-.1667em\lower.7ex\hbox{E}\kern-.125emX}}

\ifCLASSINFOpdf
\else
\fi

\ifCLASSOPTIONcompsoc
 \usepackage[caption=false,font=normalsize,labelfont=sf,textfont=sf]{subfig}
\else
 \usepackage[caption=false,font=footnotesize]{subfig}
\fi

\begin{document}

\title{End-to-end Online Speaker Diarization \\with Target Speaker Tracking}

\author{Weiqing Wang, Ming Li, \IEEEmembership{Senior Member, IEEE}
\IEEEcompsocitemizethanks{\IEEEcompsocthanksitem
Weiqing Wang and Ming Li are with the Department of Electrical and Computer Engineering at Duke University. Ming Li is also with Data Science Research Center at Duke Kunshan University.

Corresponding author:
Ming Li, E-mail: ming.li369@duke.edu
}
%\thanks{Manuscript received; revised.}
}

% Idea
% 1. Online TS-VAD
% 2. Including ResNet and ECAPA-TDNN front-end, test different segment size and shift, also the sentence length and shift
% 3. Multi-channel
% 4. Train with embedding loss
% 5. Do not use mean, use the statistical pooling
% 6. Various speakers
% 7. use WavLM as front end

\maketitle

\begin{abstract}
This paper proposes an online target speaker voice activity detection system for speaker diarization tasks, which does not require a priori knowledge from the clustering-based diarization system to obtain the target speaker embeddings. By adapting the conventional target speaker voice activity detection for real-time operation, this framework can identify speaker activities using self-generated embeddings, resulting in consistent performance without permutation inconsistencies in the inference phase. During the inference process, we employ a front-end model to extract the frame-level speaker embeddings for each coming block of a signal. Next, we predict the detection state of each speaker based on these frame-level speaker embeddings and the previously estimated target speaker embedding. Then, the target speaker embeddings are updated by aggregating these frame-level speaker embeddings according to the predictions in the current block. Our model predicts the results for each block and updates the target speakers' embeddings until reaching the end of the signal. Experimental results show that the proposed method outperforms the offline clustering-based diarization system on the DIHARD III and AliMeeting datasets. The proposed method is further extended to multi-channel data, which achieves similar performance with the state-of-the-art offline diarization systems. 
\end{abstract}

\begin{IEEEkeywords}
Speaker diarization, Online speaker diarization, end-to-end diarization, target speaker voice activity detection
\end{IEEEkeywords}

% \fi
\IEEEpeerreviewmaketitle

\section{Introduction}
\IEEEPARstart{S}{peaker diarization} is the task that assigns speaker labels to speech regions, serving as an important pre-processing pipeline for many downstream tasks like multi-speaker automatic speech recognition~\cite{ASR1, ASR2} in different scenarios, e.g., broadcast news~\cite{broadcast1, broadcast2, broadcast3}, meeting conversations~\cite{meeting1, meeting3}, conversational telephone speech~\cite{CTS1, CTS2} and clinical diagnosis~\cite{lahiri2020learning, bone16_interspeech}, etc.

Early research of speaker diarization primarily focused on modularized systems, wherein audio signals are processed through a series of cascade modules \cite{park2022review}.  With the advancement of speaker representation techniques \cite{embd1, embd2, embd3}, these modularized diarization systems demonstrated the potential to deliver commendable performance across various scenarios \cite{dihard2vbx, dihard2qingjian}. Nevertheless, these systems usually rely on clustering-based algorithms, which typically allocated audio samples to singular speaker clusters. This approach posed limitations when confronted with overlapping speech regions, as it struggled to handle situations when multiple speakers' voices are present within the same segment. Addressing this issue needs some additional modules, such as overlap speech detection \cite{boakye2008overlapped} and target-speaker voice activity detection (TS-VAD) \cite{TSVAD}. In addition to overlapping speech, each module in the modularized systems is optimized independently, which makes it difficult to optimize the whole system. 

Considering the inherent redundancy and the challenge in addressing overlapping speech within modularized systems, an framework called end-to-end neural diarization (EEND) was proposed \cite{EENDLSTM, EENDATT}. EEND aimed to streamline the entirety of the speaker diarization process by a single neural network. It can not only effectively manage overlapping speech but also directly optimize the diarization objective. Additionally, EEND exhibits an inherent adaptability, making it easily transferable to diverse domains through fine-tuning using real-world data. However, EEND also has certain limitations. For instance, it operates with a fixed maximum number of speakers. To solve this, a subsequent advancement emerged in the form of an extended EEND model incorporating an encoder-decoder-based attractor \cite{EENDEDA}. This extension was specifically designed to address scenarios with various speakers, thereby further expanding the applicability of the EEND framework.

 Actually, the aforementioned diarization systems, both the modularized one and the end-to-end one, show promising performance on many datasets, but they are both designed to process pre-recorded audio and operate offline, meaning that they can analyze the entire audio file at once to identify the individual speakers. However, there are several practical applications, such as meeting transcription systems or smart agents, where the demand for low latency in speaker assignment is paramount \cite{park2022review}. Despite ongoing efforts to develop online speaker diarization systems, both within the realm of clustering-based approaches \cite{dimitriadis2017developing,zhang22_odyssey} and neural network-based diarization frameworks \cite{zhang2019fully,xue2021online,han2021bw}, this remains a challenge that has yet to be fully overcome.

In an online modularized system, each module needs to operate in an online manner, and the key point is how to replace the offline clustering algorithm with an online one to process the speaker representations in chronological order \cite{zhu2016online, zhang22_odyssey}. Under this consideration, most of the online modularized systems only focus on the clustering algorithm and ignore other modules, e.g., overlap speech detection.

For the online EEND framework, the main challenge is the inconsistency of the speaker order accross different timestamp, as the order of speaker can be one of the permutations of all existing speakers. During training stage, this inconsistency can be solved using a permutation free objective \cite{EENDLSTM, kolbaek2017multitalker}, but it limits the EEND to be extended for online purpose as the order of speakers are always changing as a new signal coming. One solution is to employ a buffer to store both prior inputs and results to align the current results with previous one \cite{xue2021online, xue21d_interspeech, OnlineEENDGLA}, but this usually requires a large buffer for a satisfying performance.

As a post-processing method for offline modularized diarization system, target-speaker voice activity detection (TS-VAD) is usually employed to refine the diarization results from clustering algorithm \cite{TSVAD}. Given the target speaker representations and audio signals, TS-VAD can recognize the overlapping speech in a block-wise manner. This inherent property implies that TS-VAD can naturally be adapted to an online module. However, this characteristic is usually ignored as TS-VAD serves as a sub-module for the modularized diarization system. If we can obtain the target speaker representations in an online manner, TS-VAD can be easily executed in real-time. Furthermore, if the target speaker representations are generated by TS-VAD itself, it can be considered as a fully end-to-end online module. 

In this paper, we propose a fully end-to-end online framework for speaker diarization task by adapting the conventional TS-VAD for real-time operation, termed as online TS-VAD (OTS-VAD). Similar to offline TS-VAD, the OTS-VAD can not only retrieve the speaker activities for each speakers but also detect the overlapping speech. Unlike the EEND framework, our proposed system determines speaker activities using target speaker embeddings. These embeddings are self-generated by the OTS-VAD model and continually updated with incoming audio signals. As the order of these target speaker embeddings remains constant, inconsistencies during the inference phase are eliminated. Therefore, our approach can employ a small block size during the inference stage, ensuring reduced memory consumption.

This paper builds upon our earlier research \cite{OnlineTSVAD} focused on online speaker diarization. The new contributions of this extension include:

\begin{itemize}
\item Improving the performance to state-of-the-art (SOTA) levels through refined training and inference methodologies.
\item Extending the model to accommodate multi-channel data, ensuring improved and consistent performance.
\item Replacing the front-end ResNet model with a lightweight Conformer, which indicates the possibility of reducing the model's size without significant performance trade-offs.
\end{itemize}

The organization of this paper is as follows. Section II reviews the offline and online methods for both modularized system and end-to-end frameworks. Section III presents the proposed online TS-VAD method as well as the extensions. Section IV describes the experimental details and Section V presents the results, respectively. Section VI provides the conclusions. 

\section{Related Works}
\subsection{Modularized speaker diarization system}
\subsubsection{Offline method}
The modularized speaker diarization system, often referred to as the traditional speaker diarization system, is composed of several sub-modules.  The process begins with voice activity detection (VAD) that filters out non-speech components like background noise from the speech signal \cite{ryant2013speech, thomas2014analyzing, gelly2017optimization}. Following this, the audio stream is segmented into speaker-consistent sections \cite{senoussaoui2013study, sell2018diarization}. From these segments, speaker embeddings are derived \cite{ivector, embd1}. Leveraging a suitable similarity measurement technique, clustering algorithms such as agglomerative hierarchical clustering (AHC) \cite{diez2020optimizing} or spectral clustering \cite{von2007tutorial} are utilized to group these embeddings. Sometime the pre- and post-processing modules are also included for better performance, such as speech enhancement \cite{erdogan2015phase}, speech separation \cite{xiao2021microsoft}, overlapping speech detection \cite{boakye2008overlapped} and resegmentation \cite{diez2019analysis}, etc. In addition, two or more of the modules can be combined by using a single neural network, e.g., the region proposal network is employed to perform VAD, segmentation and embedding extraction at a time \cite{huang2020speaker}. 

\subsubsection{Online method}
While VAD, segmentation, and speaker embedding can be executed in real-time, the crux of transitioning the modularized system to an online format lies in substituting the offline clustering technique with an online version. Notable approaches include the adapted i-vector using a transformation matrix \cite{zhu2016online} and modified clustering methods \cite{dimitriadis2017developing}. However, these approaches can be inefficient for extended audio clips due to their time complexities scaling linearly with the number of speaker segments. Other online clustering methods, like incremental clustering \cite{coria2021overlap} and online spectral clustering \cite{xia2022turn} were also proposed for online diarization task with low latency. In addition, several supervised methods such as UIS-RNN \cite{zhang2019fully} and UIS-RNN-SML \cite{fini2020supervised} were proposed, where the segmentation and clustering methods are replaced with a trainable model. 

\subsection{End-to-end speaker diarization system}
\subsubsection{Offline method}
End-to-end neural diarization (EEND) is an innovative framework that performs the complete speaker diarization process using a single deep neural network \cite{EENDLSTM, EENDATT}. Given a speech signal, it outputs the diarization results directly, as well as recognizing overlapping speech. To enhance its versatility and tackle scenarios with an undefined number of speakers, an improved version known as EEND with Encoder-Decoder-Based Attractor (EDA) was introduced. This advancement not only bolstered performance but also widened the framework's utility. Further refinements include integrating front-end components like Conformer \cite{liu21j_interspeech} and introducing EEND with Global and Local Attractors (GLA) \cite{horiguchi2021towards}, both of which have enhanced the system's efficacy on real datasets. In addition to the fully end-to-end framework, EEND can also be employed in modularized speaker diarization system, such as embedding extraction \cite{kinoshita2021integrating,kinoshita21_interspeech,kinoshita2022tight}, overlap-aware resegmentation \cite{bredin21_interspeech} and post-processing \cite{horiguchi2021end}.

\subsubsection{Online method}
The EEND framework has also been widely explored in online speaker diarization task. Han et al. \cite{han2021bw} proposed a block-wise EEND-EDA (BW-EDA-EEND) which uses hidden states in previous blocks to generate attractors in a sequential manner. Another approach to tackle the issue of speaker permutation ambiguity involves maintaining acoustic features within a speaker-tracing buffer \cite{xue2021online, xue21d_interspeech}.  On top of this approach, Horiguchi et al. \cite{OnlineEENDGLA} enhanced EEND-GLA for online usage by integrating a speaker-tracing buffer. Remarkably, this method achieved SOTA performance across various datasets, outperforming numerous offline modularized speaker diarization systems.

\subsection{Target speaker voice activity detection}
Target speaker tracking is employed in many speech-related tasks to retrieve the information of a specific speaker, including target speaker automatic speech recognition (TS-ASR) \cite{TSASR}, target speaker speech separation \cite{li20p_interspeech} and TS-VAD \cite{TSVAD}. The similarity of these tasks is that they all use a speaker profile to focus on the speech of interest, which consequently refines results corresponding to that particular speaker. Impressively, this method remains effective even when the audio encompasses overlapping speech. Typically, the speaker profile constitutes a pre-enrolled speaker embedding that contains the speaker identity information like i-vector \cite{ivector} and x-vector \cite{embd1}.

For speaker diarization, TS-VAD has been introduced to refine the results derived from clustering-based diarization systems. It is particularly useful in situations where there are many overlapping speech in a recording, and the focus is on detecting the presence or absence of a specific speaker. The background of TS-VAD can be traced back to personal VAD \cite{ding20_odyssey}, which adapt the VAD process to the unique vocal characteristics of an individual speaker. However, personal VAD method usually does not have good performance in diarization task as it only consider a single speaker and ignore the relationship between speakers in the speech signal. TS-VAD, in contrast, is designed to simultaneously recognize speech activities of multiple speakers present at any given segment. Crucially, it incorporates the constraints and relationships among different speakers, ensuring a more comprehensive understanding. As a result, TS-VAD offers an enhanced performance and delivers more reliable diarization outcomes.

\begin{figure}[b]
\centering
\includegraphics[scale=1.5]{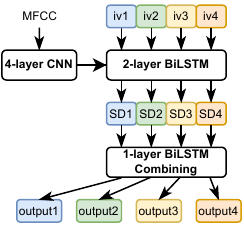}
\caption{The architecture of the i-vector-based TS-VAD \cite{TSVAD}}.
\label{fig:TS-VAD-iv}
\end{figure}

The early version of TS-VAD employs a pre-enrolled i-vector as acoustic footprint \cite{TSVAD}. It takes a short speech segment and several i-vectors as input and evaluate if one of the frames belongs to one or more speakers, as shown in Fig. \ref{fig:TS-VAD-iv}. Fig. \ref{fig:TS-VAD-iv} shows an i-vector-based TS-VAD model. First, a CNN takes the acoustic features as input and produce a temporal feature map. Next, each i-vector is concatenated with each frame of the feature map, and they are separately fed to the 2-layer BiLSTM, producing a speaker detection (SD) feature for each i-vector. These SD features are concatenated again as the input of the final BiLSTM layer and converted to the posterior presence probabilities of each speaker. Remarkably, the i-vector-based TS-VAD approach demonstrates a notable advantage in terms of diarization error rate (DER) when pitted against traditional clustering-based methods, which brings forth a robust solution for the diarization challenge.

The deployment of i-vectors in TS-VAD, while successful to some extent, showed limitations when applied to multi-scenario data \cite{wang21y_interspeech}. These findings pave the way for the exploration of x-vectors as an alternative, but a simple swap of i-vectors for x-vectors did not yield an immediate boost in performance \cite{TSVAD}. The speculated cause of this unexpected behavior is that the shallow CNN employed in the front-end of the TS-VAD system might not be adept at efficiently processing the deep learning-driven x-vector. To address this conjecture, Wang et al. \cite{wang2022similarity} replaced the front-end of the TS-VAD with a pre-trained front-end module tailored for x-vectors. This modified architecture demonstrated a more competent handling of x-vectors, consequently leading to a superior performance when benchmarked against the i-vector-based system. Additionally, the intrinsic robustness and generalization capability of x-vectors empowered this model to perform commendably on challenging datasets like DIHARD III, which represents diverse real-world scenarios.

Usually, TS-VAD can only handle fixed number of speakers, as the final layer is a linear projection in general. A common strategy is to define the output size based on the maximum anticipated number of speakers. This presents a challenge because, in real-world scenarios, the actual number of speakers is indeterminate. Furthermore, when the input audio features fewer speakers than the maximum set, the TS-VAD model performs redundant computations, especially when certain speaker embeddings are initialized to zero or arbitrary vectors. To address this inefficiency, the terminal linear layer can be substituted with a recurrent neural network, capable of managing variable-length data. For instance, both LSTM \cite{cheng2023multi} and Transformer \cite{wang2023target} have been implemented along the speaker dimension, rather than the temporal one. Recently, Cheng et al.~\cite{cheng2023target} proposed a sequence-to-sequence (Seq2Seq) architecture for TS-VAD, termed as Seq2Seq TS-VAD. In this approach, frame-level representations and speaker embeddings are input independently into the encoder and decoder, obviating the need for concatenation. Notably, the decoder creates an embedding of consistent dimensionality for each speaker's voice activity, independent of the duration of the input features. Additionally, the final linear layer offers enhanced flexibility, facilitating voice activity predictions at finer temporal resolutions with modest computational demands.

Compared to the end-to-end neural diarization (EEND), TS-VAD serves as a post-processing module within the modularized speaker diarization system. While EEND offers an all-encompassing solution, learning speaker embeddings directly from audio waveforms to produce diarization outcomes, TS-VAD refines the outputs of clustering-based methods. Owing to its reliance on prior knowledge derived from clustering-based diarization results, TS-VAD typically exhibits superior performance to EEND. In the subsequent section, we introduce our TS-VAD-based end-to-end online framework, wherein the speaker embedding is self-generated by the TS-VAD model.

\begin{figure*}[!htbp]
\centering
\includegraphics[scale=1.2]{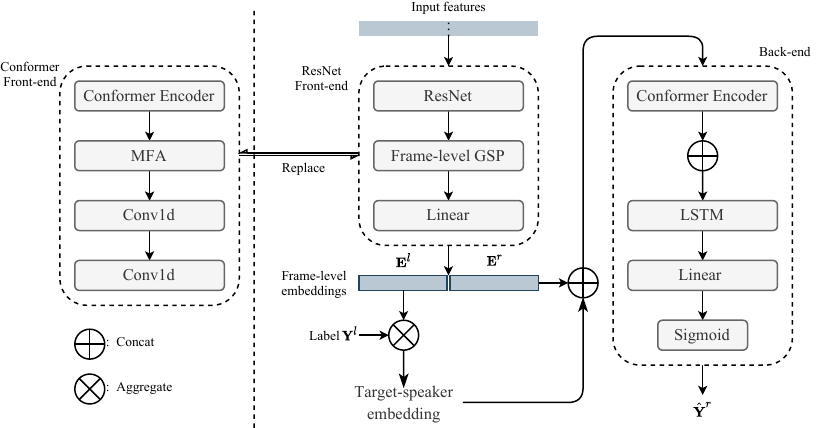}
\caption{The architecture of the proposed OTS-VAD. The rigth is the backbone with ResNet-based front-end, and the left is another Conformer-based front-end which can replace the ResNet-based front-end.}
\label{fig:TS-VAD-ResNet}
\end{figure*}

\section{Online Target Speaker Voice Activity Detection}
\label{sec:ts-vad}

This section elucidates the workings of the x-vector-based TS-VAD and details its transformation from an offline to an online approach, encompassing both the training and inference procedures. 

\subsection{X-vector-based TS-VAD}
Consider $N$ target speaker embeddings represented as $\mathbf{E} = \{\mathbf{e}_1, \mathbf{e}_2, ..., \mathbf{e}_N\}$, and the acoustic feature as $\mathbf{X}=\{\mathbf{x}_1, \mathbf{x}_2, ..., \mathbf{x}_T\}$. A typical TS-VAD model detects the speech activities of each speaker \( n \) as: 

\begin{equation}
    \label{eq:TSVAD1}
    \mathbf{S}^{'}_n = [\mathbf{S}^{'}_{n,1}, \mathbf{S}^{'}_{n,2}, ... , \mathbf{S}^{'}_{n,T}] = f_d(
    \left[\begin{matrix}\mathbf{e}_n\\\hat{\mathbf{e}}_1\end{matrix}\right],
    \left[\begin{matrix}\mathbf{e}_n\\\hat{\mathbf{e}}_2\end{matrix}\right], ...,
    \left[\begin{matrix}\mathbf{e}_n\\\hat{\mathbf{e}}_T\end{matrix}\right]),
\end{equation}
where $f_d$ is the detection function that produces the speaker existence probability $\mathbf{S}^{'}_{n,t}\in R$ based on target speaker embedding $\mathbf{e}_n$ at time t, and $\hat{\mathbf{e}}_t$ is the frame-level embedding generated from the acoustic features by the front-end module. Subsequently, the speech activities from all speakers are concatenated, serving as the last layer's input. This generates refined speech activities \( \mathbf{S}\in (0,1)^{N \times T} \) for all speakers:

\begin{equation}
    \label{eq:TSVAD2}
    \mathbf{S} = f_j(\left[\begin{matrix}\mathbf{S}^{'}_{1,1}& ...& \mathbf{S}^{'}_{1,T}\\...& \mathbf{S}^{'}_{n,t}& ...\\\mathbf{S}^{'}_{N,1}&  ...& \mathbf{S}^{'}_{N,T}\end{matrix}\right]),
\end{equation}
where \( f_j \) acts as the joint detection function that produces final speech activities. As depicted in Fig. \ref{fig:TS-VAD-iv}, \( f_d \) operates as the 2-layer BiLSTM, while \( f_j \) functions as the 1-layer BiLSTM. The front-end model designed for frame-level embedding extraction is constituted by a 4-layer CNN, and this is the typical i-vector TS-VAD framework. 

However, this framework does not work well with x-vector, even the performance of x-vector is better than the i-vector \cite{TSVAD}. As suggested in \cite{wang2022similarity}, we can replace the front-end module with a pre-trained network that is used for target speaker embedding extraction, which makes the frame-level embedding learn faster to capture the identity information in target speaker embedding. This x-vector-based TS-VAD method achieved the state-of-the-art performance on DIHARD III dataset and was further improved by the x-vector-based sequence-to-sequence TS-VAD framework \cite{cheng2023target}.

\subsection{Online TS-VAD}
\label{sec:method}

\subsubsection{Architecture}

Fig. \ref{fig:TS-VAD-ResNet} shows the architecture of the ResNet-based online TS-VAD (OTS-VAD). The model contains two sub-modules: front-end for embedding extraction and back-end for target speaker detection. The front-end consists of a ResNet and a fully-connected layer following a frame-level global statistics pooling (GSP) summarizes the statistics of each frame of the ResNet output. The back-end contains an conformer-based encoder, a BiLSTM layer and a linear layer.

% Given an acoustic features $\mathbf{X}\in\mathbb{R}^{H \times L}$, the ResNet extracts a temporal feature map $\mathbf{M} \in \mathbb{R}^{C \times \frac{H}{8} \times \frac{L}{8}}$, where $C=256$ is the number of channels, H is the dimension of features, and L is the number of frames. The temporal resolution of the feature map is eight times less than that for the input features as ResNet can downsample the input by eight times. Next, the frame-level GSP layer extracts the mean and standard deviation over feature dimension, producing a $2C$-dim vector for each frame. Finally, the fully-connected layer takes the GSP output $\mathbf{P}\in\mathbb{R}^{T \times 2C}$ and extracts the frame-level embeddings $\hat{\mathbf{E}}=\left[\hat{\mathbf{e}}_1, ..., \hat{\mathbf{e}}_T\right] \in \mathbb{R}^{T \times D}$, where $T=\frac{L}{8}$ and $D$ is the size of frame-level embeddings. $D$ is also the size of the target speaker embeddings. The usage of frame-level GSP is to imitate the utterance-level GSP in the speaker recognition task \cite{cai2020within}, where we consider each frame as an utterance and extract the embeddings. In this way, the front-end module can capture the speaker identity information in each frame as the x-vector does, leading to a better performance. 

Given an acoustic feature $\mathbf{X}\in\mathbb{R}^{H \times L}$, the ResNet transforms it into a temporal feature map $\mathbf{M} \in \mathbb{R}^{C \times \frac{H}{8} \times \frac{L}{8}}$. Here, $C$ represents the number of channels, $H$ is the feature's dimensionality, and $L$ indicates the total number of frames. It is worth noting that the temporal resolution of this feature map is reduced by a factor of eight compared to the original input, attributed to ResNet's capability of downsampling by this magnitude. Subsequently, the frame-level GSP layer computes the mean and standard deviation over the feature dimension, thus generating a vector of dimension $2C$ for each frame. The succeeding fully-connected layer ingests the GSP's output, $\mathbf{P}\in\mathbb{R}^{T \times 2C}$, producing frame-level embeddings, represented as $\hat{\mathbf{E}}=\left[\hat{\mathbf{e}}_1, ..., \hat{\mathbf{e}}_T\right] \in \mathbb{R}^{T \times D}$. Here, $T=\frac{L}{8}$ and $D$ is the dimensionality of frame-level embeddings. Furthermore, $D$ matches the dimension of the target speaker embeddings. The usage of frame-level GSP is to imitate the utterance-level GSP in the speaker recognition task \cite{cai2020within}. This method interprets each window as a distinct utterance for which embeddings are extracted. By adopting this approach, the front-end module can effectively discern the speaker's identity for each frame, which invariably leads to enhanced performance.

% After we obtain the frame-level embeddings $\hat{\mathbf{E}}$ and the target speaker embedding $\mathbf{E}=\left[\mathbf{e}_1, ..., \mathbf{e}_N\right] \in \mathbb{R}^{N \times D}$, the back-end module can predict the speech activities as offline TS-VAD does. To be more specific, each of the target speaker embedding will be repeated and combined with the frame-level embeddings. The first Encoder separately processes the concatenated embeddings for each target speaker to obtain the first level of decision, which does not necessarily represent the probabilities. Next, the decisions of all speakers are concatenated and fed to the BiLSTM for refinement, which can exchange the cross-speaker information, Finally, a fully-connected layer with sigmoid function predicts the speaker existence probabilities. 

% Therefore, the key point is to obtain the target speaker embedding without an additional speaker embedding extractor in both training and inferring process. In addition, how to keep the consistency of the target speaker embedding is also important for back-end module to detect the speech activities. In the following sections, we will describe how to obtain the consistent target speaker embedding in training and inferring process. 

Upon obtaining the frame-level embeddings $\hat{\mathbf{E}}$ and the target speaker embeddings, represented by $\mathbf{E}=\left[\mathbf{e}_1, ..., \mathbf{e}_N\right] \in \mathbb{R}^{N \times D}$, the back-end module predicts speech activities in a manner akin to the offline TS-VAD model. To be more specific, each target speaker embedding undergoes replication and is subsequently concatenated with the frame-level embeddings. The initial Encoder separately manages the concatenated embeddings corresponding to each target speaker, yielding a primary decision level. It's noteworthy that these decisions may not strictly indicate probabilities. Following this, decisions pertaining to all speakers are concatenated and fed to the BiLSTM for refinement. This stage facilitates the exchange of inter-speaker information. Finally, a sigmoid-activated fully-connected layer estimates the probabilities of speaker existence.

To make the offline TS-VAD operate in an online manner, a key point is to obtain the target speaker embedding without an additional speaker embedding extractor in both training and inferring process. In addition, how to keep the consistency of the target speaker embedding is also important for back-end module to detect the speech activities. In the following sections, we will describe how to obtain the consistent target speaker embedding in training and inferring process. 

\subsubsection{Training process}
\begin{figure}[!htbp]
\centering
\includegraphics[scale=1.4]{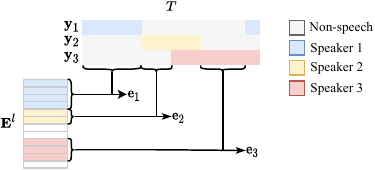}
\caption{Illustration of the target speaker embedding extraction.}
\label{fig:target-embedding}
\end{figure}

As shown in Fig. \ref{fig:TS-VAD-ResNet}, given an input $\mathbf{X}$ with length of $2L$, the initial step involves partitioning the input features uniformly into left and right segments $\mathbf{X}^l$ and $\mathbf{X}^r$, each having a length of $L$. After extracting the frame-level embeddings $\hat{\mathbf{E}}^{l}, \hat{\mathbf{E}}^{r} \in \mathbb{R}^{T \times D}$ for the left and right segments, the target speaker embedding for the $n$-th speaker can be obtained by:

\begin{equation}
    \label{eq:TS-embd}
    \mathbf{e}_n = \frac{\bar{\mathbf{y}}^l_n\hat{\mathbf{E}}^{l}}{\sum_{i=1}^T\bar{\mathbf{y}}^l_{n,i}},
\end{equation}
where $n\in\{1, 2, ... N\}$, $i$ is the frame index, $T=\frac{L}{8}$ by downsampling from ResNet, $\bar{\mathbf{y}}^l_n \in \{0, 1\}^T$ is the corresponding label of $n$-th speaker in the left segment with the overlapping regions set to zero. Eq. \ref{eq:TS-embd} describes that target-speaker embedding can be obtained by averaging all non-overlapping frame-level embeddings for each respective speaker, as shown in Fig \ref{fig:target-embedding}. In addition, due to the limit of the GPU memory, we cannot train the model with long-duration inputs. However, this will lead to a problem that $\mathbf{X}^l$ only contains limited speakers with a short duration, and the back-end module only accepts insufficient target speaker embeddings during the training stage. Therefore, the model cannot deal with the input with lots of speakers. To cope with this, we replace the $\mathbf{X}^l$ with a randomly simulated signal that contains multiple speakers, and the probability of replacement is set to 0.5. 

%After extracting the target speaker embeddings, the target-speaker embedding for each speaker can be randomly selected from $\mathbf{E}^1\sim\mathbf{E}^K$, where $\mathbf{E}^K\in\mathbb{R}^{N \times D}$. For different block, we select different target speaker embedding sets to increase the variety of training data. In addition to current K blocks, we also select some target speakers from other blocks for negative sampling. In each training batch with batch size of $H$ and input length of $T$, each training sample is split into K blocks, thus the actual batch size is $H \times K$ with input length of $\frac{T}{K}$.

After obtaining the target speaker embedding, the training process of back-end is the same as the offline TS-VAD, where we repeat and concatenate the target speaker embeddings to the frame-level embeddings as the input of back-end. Since we already know the order of the speakers in the ground truth label $\mathbf{Y}\in\{0, 1\}^{T \times N}$, we will keep the order of all speaker in the target speaker embeddings. Therefore, the speaker order in the output $\hat{\mathbf{Y}}\in(0, 1)^{T \times N}$ is the same as the labels and we do not need to solve the permutation ambiguity as presented in the EEND framework. Then we optimized the whole model by the binary cross entropy (BCE) loss.

It's important to mention that we exclude silent regions from the input audio beforehand. This is because when there's an abundance of silence in the data, the aggregation method may struggle to gather sufficient target speaker embeddings within a relatively short segment for training purposes. %Therefore, this makes it difficult for the model to learn the relation between the target speaker embeddings and the frame-level embeddings. 

\begin{figure}[!htbp]
\centering
\includegraphics[scale=1.2]{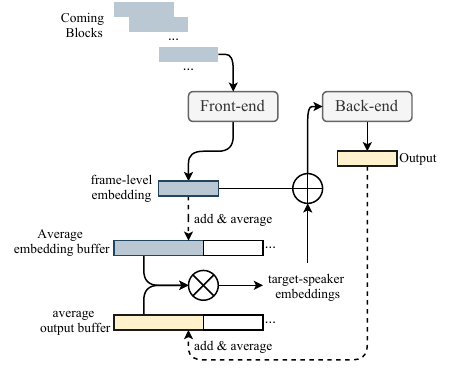}
\caption{Inference process of OTS-VAD with embedding buffer and output buffer. Dash line means updating the buffer after obtaining the output from current block.}
\label{fig:inference1}
\end{figure}

\subsubsection{Inference process}
The inference process is quite different from the training process as the target speaker embeddings are accumulated during the inference process. Specifically, during inference, target speaker embeddings are aggregated incrementally as the OTS-VAD system systematically processes each successive block of audio. Here we also assume that the silence regions are removed in advance. For clarity, we assume that the features and output have the same time resolution, and we will explain the actual time resolution for all features and output in Section \ref{sec:exp}.

As shown in Fig. \ref{fig:inference1}, the length and shift for each coming block is $l$ and $m$, respectively. In addition, we need a frame-level embedding buffer $\hat{\mathbf{E}}\in\mathbb{R}^{T \times D}$and a output buffer $\hat{\mathbf{Y}}\in(0, 1)^{T \times N}$ to store the frame-level embedding and corresponding output over time, where $T$ is the total length of output, $D$ is the embedding size and $N$ is the preset number of speakers. These two buffers are both initialized with zeros. Initially, the block size begins at $m$ and expands until there is a sufficient signal to compose a full block of size $l$. Following this, the block size remains constant at $l$, advancing with a consistent shift of $m$. Notably, $m$ also designates the latency of the OTS-VAD system.

For the first block, we believe that it only contains one speaker as $m$ is set to a very small value. This assumption makes it very easy to initialize the system using the first block. The first block is fed into front-end module, producing the initialized frame-level embeddings. In addition, we assign the output of the first speaker with one, and other values in the output are set to zero:
\begin{equation}
    \hat{\mathbf{y}}^1_{n,t} =     \left\{ \begin{array}{rcl}
         1 & , & n=1\\
         0 & , & n=2, 3, ..., N
                \end{array}\right.
\end{equation}
where $\hat{\mathbf{y}}^1$ is the output of the first block, n and t are the index of speaker and frame, respectively. Finally, both the frame-level embeddings and the output are added to both buffer at the corresponding position and averaged at the overlapping region between the blocks, as shown in Fig. \ref{fig:inference1}.

For the following blocks, assume that we already have the averaged embeddings buffer $\hat{\mathbf{E}}$ and output buffer $\hat{\mathbf{Y}}$ until timestamp $t^{'}$, the $n$-th target speaker embedding can be easily computed by using the first $t^{'}$ frames of these two buffers:
\begin{equation}
    \label{eq:TS-embd-infer}
    \mathbf{e}_n = \frac{\bar{\mathbf{Y}}_n\hat{\mathbf{E}}}{\sum_{t=1}^{t^{'}}\bar{\mathbf{Y}}_{n,t}},
\end{equation}
where $\bar{\mathbf{Y}} \in \{0, 1\}^{l \times N}$ is the output binarized from $\hat{\mathbf{Y}}$ with an \textbf{upper threshold} $thres_\text{upper}$. Here the upper threshold is usually greater than 0.5, which means that we only use the frame-level embedding with higher speaker existence probabilities for target speaker embedding extraction to ensure the purity of embeddings. Given the target speaker embeddings and frame-level embeddings, the back-end module can easily predict the output for current block. If a new speaker presents in the last $m$ frames without overlapping speech, the OTS-VAD cannot recognize any speaker given previous target speaker embedding, thus all the values of the output are closed to zero in the last $m$ frames. In this way, we can know that a new speaker appears, and we can assign the output with the value of the upper threshold for the new speaker in the last $m$ frames. If the number of speakers has already reached the maximum, we keep the output unchanged. Usually, if we find that all values of the output in the last $m$ frames are lower than a \textbf{lower threshold} $thres_\text{lower}$, we believe that there is a new speaker, where the lower threshold is set to no more than 0.5.

An important thing for an online application is the latency. If there comes unlimited number of blocks, it is impossible to store all frame-level embeddings in the buffers, and the computation cost also becomes unacceptable for calculating the target-speaker embedding. In that case, we can only store those frames-level embeddings with the higher probabilities for each speaker in the buffer and discard others, which significantly reduces the computation cost when the audio signal is extremely long. 

\begin{figure}[!htbp]
\centering
\includegraphics[scale=1.4]{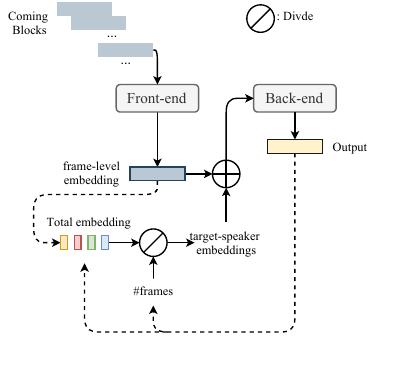}
\caption{Inference process of OTS-VAD with accumulated embedding and the total number of frames. Dash line means updating the accumulated embedding and the number of frames after obtaining the output from current block.}
\label{fig:inference2}
\end{figure}

Another solution for reducing the latency is to accumulate the frame-level embedding for each speaker and record the sum of all frame-level embeddings, referred to accumulated total embedding $\tilde{\mathbf{e}}_n$, as shown in Fig. \ref{fig:inference2}. At the same time, the number of frames $F$ is also recorded. Then the target speaker embedding $\mathbf{e}_n$ can be directly computed by:
\begin{equation}
    \mathbf{e}^{k+1}_n = \frac{\tilde{\mathbf{e}}^k_n}{F^k},
\end{equation}
where $k$ is the index of block and $\mathbf{e}^k_n$ is the target speaker embedding for inference of the $k$-th block. When there comes a series of frame-level embeddings $\hat{\mathbf{E}}^k = \left[\hat{\mathbf{e}}_1, ..., \hat{\mathbf{e}}_T\right]\in\mathbb{R}^{T \times D}$ and the corresponding binarized prediction $\bar{\mathbf{Y}}^k\in\{0, 1\}^{N \times T}$ from the $k$-th block, the accumulated embedding and number of frame for $n$-th speaker can be updated by:
\begin{equation}
\begin{split}
    \tilde{\mathbf{e}}^{k}_n &= \tilde{\mathbf{e}}^{k-1}_n + \bar{\mathbf{Y}}^k_n\hat{\mathbf{E}}^k, \\
    F^k &= F^{k-1} + \sum_{i=0}^T\bar{\mathbf{Y}}^k_{n, t}.
\end{split}
\end{equation}
Similar to the buffer-based inference process mentioned in Fig. \ref{fig:inference1}, only the frame with the probabilities greater than the upper threshold will be used for updating. If the probabilities in the last $M$ frames are lower than the lower threshold, we notify a new speaker coming by setting his/her probabilities to upper threshold.

The first strategy utilizes a buffer to continuously store average frame-level embeddings and corresponding outputs. This method, by retaining a comprehensive record of embeddings, ensures greater precision in target speaker inference, thereby potentially enhancing system performance. However, this method's primary limitation emerges when processing extensive input features, resulting in increased computational time and memory requirements. On the other hand, the second approach streamlines the inference process by only calculating the accumulated embeddings and their corresponding frame counts. While this method offers significant computational efficiencies, especially in scenarios with prolonged input sequences, there is a potential slight degradation in performance accuracy. Consequently, the optimal choice between these strategies would depend on the specific application requirements, balancing between accuracy and computational resource constraints.

\subsection{Incorporation of MFA-Conformer as front-end}

Multi-scale feature aggregation (MFA) is a method that joins together output feature maps from every frame-level module within a speaker embedding network before pooling at the utterance-level. Research indicates that when used with TDNN-based networks, this method enhances performance, which suggests that features from the lower-levels offer valuable speaker-related information \cite{embd3}.

To incorporate the Conformer encoder into speaker verification tasks, the MFA-Conformer introduces an MFA module within the Conformer encoder \cite{zhang22h_interspeech}. This module combines all the frame-level outputs from the Conformer blocks before moving them to the pooling layer. Once this combined frame-level feature map is obtained, attentive statistics pooling is used to generate an utterance-level representation \cite{okabe18_interspeech}. Subsequently, the speaker's embedding is derived by employing batch normalization and a connected layer on this utterance-level representation. To replace the ResNet with the MFA Conformer in the OTS-VAD framework, the combined frame-level features before pooling are considered as the frame-level embeddings, which is the output of the MFA layer shown in Fig \ref{fig:TS-VAD-ResNet}. However, the time resolution of the outputs from the MFA layer is too high to be adopted as the frame-level embedding as processing this feature will consume large GPU memory in the back-end module. We then add two more 1d convolutional layers to reduce the time resolution, each of which reduces the time resolution by 2. Finally, the output of the last 1d convolutional layer has the same time resolution as the ResNet-based front-end does.

\subsection{Multi-channel extension}

Cross-channel attention mechanism has achieved significant success in multi-channel speech signal processing domains, including speech enhancement \cite{CrossChannel1}, speech separation \cite{CrossChannel2, CrossChannel3}, speech recognition \cite{CrossChannel4}, and speaker diarization \cite{CrossChannel5}. Their efficacy lies in their ability to comprehend non-linear contextual relationships across channels both temporally and contextually.

Given $C$ channels of Fbank sequence denoted by $\mathcal{X} = (\mathbf{X}^1, ..., \mathbf{X}^C)$, the target-speaker embedding from $N$ target speakers represented as ${\mathcal{E}} = ({\mathbf{E}}^{1}, ..., {\mathbf{E}}^{C})$, where ${\mathbf{E}}^{c} = [\mathbf{e}^c_1, \mathbf{e}^c_2, ..., \mathbf{e}^c_N]$ are the target speaker embeddings extracted from the c-th channel, the corresponding target speaker decision is denoted by $\mathbf{Y} = (\mathbf{y}_1, ..., \mathbf{y}_T)$. Here, $\mathbf{X}_i \in \mathbb{R}^{L \times H}$ has $L$ frames, ${\mathbf{E}}^{c}\in \mathbb{R}^{N \times D}$ signifies the D-dimensional speaker embedding from N target speakers in the c-th channel, and $\mathbf{y}_t \in \{0, 1\}^N$ encapsulates the target speaker decision at time step $t$ with a dimension of $N$. The front-end ResNet takes the $C$-channels of Fbank sequence as input, producing the frame-level speaker embedding sequence, denoted by $\hat{\mathcal{E}} = (\hat{\mathbf{E}}^{1}, ..., \hat{\mathbf{E}}^{C})$, where $\hat{\mathbf{E}}^{c} = [\hat{\mathbf{e}}^{c}_1, \hat{\mathbf{e}}^{c}_2, ..., \hat{\mathbf{e}}^{c}_T]$. Subsequent operations entail repeating the target-speaker embedding $T$ times and the frame-level speaker embedding $N$ times. Post these operations, the two speaker embeddings are concatenated in the embedding dimension, forming the matrix ${\mathcal{G}} = (\mathbf{G}^1, ..., \mathbf{G}^C)$, where ${\mathbf{G}}^{c}\in \mathbb{R}^{T \times N \times 2D}$.

The cross-channel self-attention mechanism ingests the concatenated speaker embedding $\mathbf{G}_\text{in}$:

\begin{align}
    \mathbf{G}_\text{in} &= \text{concat}(\mathcal{G}) \in \mathbb{R}^{T \times N \times C \times 2D} \\
    \mathbf{Q}^i &= \mathbf{W}_Q^i\mathbf{G}_\text{in} + \mathbf{b}_Q^i \\
    \mathbf{K}^i &= \mathbf{W}_K^i\mathbf{G}_\text{in} + \mathbf{b}_K^i \\
    \mathbf{V}^i &= \mathbf{W}_V^i\mathbf{G}_\text{in} + \mathbf{b}_V^i,
\end{align}
with $\mathbf{Q}^i$, $\mathbf{K}^i$, and $\mathbf{V}^i$ representing the query, key, and value matrices, respectively, for the $\text{i}^{\text{th}}$ attention head. The weight matrices $\mathbf{W}^i \in \mathbb{R}^{M \times 2D}$ and bias vectors $\mathbf{b}^i \in \mathbb{R}^{M}$ correspond to the $\text{i}^{\text{th}}$ head. The scaled dot-product attention mechanism is then applied:

\begin{align}
    \text{Attention}(\mathbf{Q}^i, \mathbf{K}^i, \mathbf{V}^i) = \text{softmax}\frac{\mathbf{Q}^i(\mathbf{K}^i)^\intercal}{\sqrt{n}}\mathbf{V}^i,
\end{align}
where $n=M$. Subsequently, a position-wise feed-forward layer complemented by a ReLU activation function is employed, incorporating layer normalization and residual connections at each layer. The resultant output is denoted by $\mathbf{G}_\text{out} \in \mathbb{R}^{T \times N \times C \times 2D}$. This output is averaged across the channel dimension via a global average pooling layer:

\begin{align}
\mathbf{G}^\prime = \frac{1}{C}\sum_{i=1}^{C}\mathbf{G}_{\text{out},i},
\end{align}
where $\mathbf{G}^\prime \in \mathbb{R}^{T \times N \times 2D}$. The back-end model subsequently processes this aggregated speaker embedding $\mathbf{G}^\prime$, akin to how the single-channel TS-VAD processes the concatenated speaker embedding.

\newcommand{\cmark}{\ding{51}}%
\newcommand{\xmark}{\ding{55}}%
\begin{table*}[htbp!]
  \caption{ ResNet front-end DER (\%) and JER (\%) on different datasets with the embedding buffer-based inference process in Fig. \ref{fig:inference1}. }
  \label{tab:results_resnet_infer1}
  \centering
  \begin{tabular}[c]{lccccccccccc}
    \toprule
     \multirow{4}*{\textbf{Dataset}} & \multirow{4}*{\textbf{Pretraining}} & \multirow{4}*{\textbf{Front-end}} & \multirow{4}*{\textbf{Block shift}} & \multicolumn{8}{c}{\textbf{Block length}}  \\
     \cmidrule(lr){5-12} 
     & & & & \multicolumn{2}{c}{2s} & \multicolumn{2}{c}{4s} & \multicolumn{2}{c}{8s} & \multicolumn{2}{c}{16s} \\
     \cmidrule(lr){5-6} \cmidrule(lr){7-8}  \cmidrule(lr){9-10} \cmidrule(lr){11-12}  
     & & & & DER & JER & DER & JER & DER & JER & DER & JER \\
     
   \midrule
	\multirow{7}*{{DIHARD III Eval}} & \multirow{7}*{\cmark} & \multirow{3}*{ResNet}
    &   0.4s & 16.20 & 35.50 & 14.47 & 34.16 & 13.83 & 34.57 & 13.77 & 35.09 \\
	&& & 0.8s & 15.31 & 34.65 & 13.96 & 34.05 & 13.57 & 33.87 & \textbf{13.31} & {33.83} \\
	&& & 1.6s & 16.60 & 37.00 & 14.40 & 35.17 & 14.12 & 34.76 & 13.59 & 34.34 \\
    \cmidrule(lr){3-12} && \multirow{3}*{Conformer}
	&    0.4s & 16.39 & 40.45 & 15.55 & 39.33 & 15.77 & 41.05 & 16.30 & 42.22 \\
	&&& 0.8s & 16.18 & 40.93 & 15.66 & 40.08 & \textbf{15.33} & 40.10 & 15.86 & 40.98 \\
	&&& 1.6s & 16.97 & {38.45} & 16.01 & 41.71 & 16.36 & 42.21 & 17.02 & 42.82 \\
   \midrule
	\multirow{7}*{{DIHARD III Eval}} & \multirow{7}*{\xmark} & \multirow{3}*{ResNet}
	&  0.4s & 19.83 & 44.34 & 16.86 & 40.13 & 15.58 & 39.12 & 16.02 & 39.91 \\
	&&& 0.8s & 20.25 & 44.06 & 17.06 & 40.32 & 15.35 & 38.43 & 15.48 & 39.23 \\
	&&& 1.6s & 21.10 & 43.58 & 16.82 & 40.55 & 15.51 & 39.23 & \textbf{14.67} & {38.46} \\
    \cmidrule(lr){3-12} && \multirow{3}*{Conformer}
    &   0.4s & 22.54 & 48.69 & 20.57 & 46.70 & 19.79 & 47.73 & 20.91 & 50.65 \\
	&&& 0.8s & 22.74 & 49.91 & 20.17 & 45.73 & 18.11 & 44.23 & 18.63 & 46.09 \\
	&&& 1.6s & 26.56 & 49.49 & 19.60 & 45.30 & \textbf{17.97} & 43.85 & 18.28 & 44.68 \\
 %   \midrule
	% \multirow{7}*{{Alimeeting Eval (1st Channel)}} & \multirow{7}*{\cmark} & \multirow{3}*{ResNet}
 %    &  0.4s &  &  &  &  &  &  &  &  \\
	% &&& 0.8s&  &  &  &  &  &  &  &  \\
	% &&& 1.6s &  &  &  &  &  &  &  &  \\
 %  \cmidrule(lr){3-12} && \multirow{3}*{Conformer}
 %    &   0.4s &  &  &  &  &  &  &  &  \\
	% &&& 0.8s &  &  &  &  &  &  &  &  \\
	% &&& 1.6s &  &  &  &  &  &  &  &  \\
 % \midrule
	% \multirow{7}*{{Alimeeting Test (1st Channel)}} & \multirow{7}*{\cmark} & \multirow{3}*{ResNet}
	% &   0.4s &  &  &  &  &  &  &  &  \\
	% &&& 0.8s &  &  &  &  &  &  &  &  \\
	% &&& 1.6s &  &  &  &  &  &  &  &  \\
 %  \cmidrule(lr){3-12} && \multirow{3}*{Conformer}
 %    &   0.4s &  &  &  &  &  &  &  &  \\
	% &&& 0.8s &  &  &  &  &  &  &  &  \\
	% &&& 1.6s &  &  &  &  &  &  &  &  \\
     \bottomrule
     \end{tabular}
\end{table*}

\section{Experimental Setup}
\label{sec:exp}
\subsection{Dataset}
{\color{black}The experiments are conducted on the DIHARD III dataset \cite{ryant2020third} and AliMeeting dataset \cite{M2MeT}. DIHARD III dataset is a multi-domain dataset that contains 34.15 hours of training data and 33.01 hours for evaluation data, including 11 different domains. The number of speakers in each recording ranges from 1 to 9. The average speech overlap ratio is 10.75\% for the development set and 9.37\% for the evaluation set, respectively. The Alimeeting dataset is an far-field 8-channel dataset that contains 118.75 hours of speech data, and it is divided into 104.75 hours for training, 4 hours for evaluation, and 10 hours for testing. The number of participants within one meeting session ranges from 2 to 4. The duration of each session is about 30 minutes. The average speech overlap ratio is 42.27\% for the training set and 34.76\% for the evaluation set, respectively. 

The existing datasets for diarization tasks generally lack sufficient complexity for effective multi-speaker learning. Therefore, data simulation becomes imperative to train a model with robust generalization capabilities. For instance, to create our training data for DIHARD III dataset, we utilize the ground truth labels from the DIHARD III development dataset. First, we derive a label matrix $\mathbf{Y}\in \{0,1\}^{N \times T}$ for each recording within the DIHARD III development dataset, where $N$ denotes the number of speakers and $T$ signifies the duration of an utterance after excluding non-speech regions. The simulated data is then generated on-the-fly in an online manner, where we randomly choose a segment of the label and fill the active regions with the continuous non-overlapped speech segments from Voxceleb dataset. Similarly, for alimeeting dataset, the data is directly simulated from the Alimeeting Train set.  

We perform data augmentation with MUSAN \cite{MUSAN} and RIRs corpus \cite{RIRS}. For the MUSAN corpus, ambient noise, music, television, and babble noise are used for the background additive noise. For the RIRs corpus, only impulse responses from small and medium rooms are employed to perform convolution with training data.

\subsection{Network Configuration}
For the ResNet-based OTS-VAD model, the front-end architecure is the same as the ResNet34 in \cite{wang2020dku}. After obtaining the temporal feature map, we apply the global statistics pooling over each frame of the feature map to produce the frame-level representation. Finally, the linear layer with 256 units projects the representation to frame-level embedding. For the back-end, it contatins a 6-layer 8-head Conformer Encoder with a 512 dimensional feed-forward layer, and the dropout is set to 0.1. The hidden size of the single BiLSTM layer is 256. The fully-connected layer projects the output from BiLSTM to a $N$-dimensional vector, where $N$ is the number of target speakers. In our experiment, $N$ is set to 8 for DIHARD III dataset and 4 for Alimeeting dataset. The total number of parameters in the ResNet-based front-end is 20.54M.

For the Conformer-based OTS-VAD model, we employ the Conformer implemented by NEMO toolkit \cite{NEMO} and we follow the speaker verification training protocal in \cite{cai2023leveraging}. The NEMO Conformer is available in three variants: small, medium, and large. For our research, we opted for the 'small' variant\footnote{\url{https://catalog.ngc.nvidia.com/orgs/nvidia/teams/nemo/models/stt_en_conformer_ctc_small}}, characterized by a convolution subsampling rate of 14 and a uniform kernel size of 31 within its convolution modules. This version encompasses 16 Conformer layers with an encoder dimension of 176, accommodates 4 attention heads, and comprises 704 linear hidden units. We only use the first 8 layers for the speaker verification pretraining. The features processed by MFA layer are then fed to two 1d convolution layers, each of which has kernel size of 3, stride of 2 and padding of 1. The back-end is the same as the ResNet-based OTS-VAD model. The total number of parameters in the Conformer-based front-end is 9.38M.

For the mutli-channel OTS-VAD, we employ the same ResNet34 as the front-end model. The back-end contains a cross-channel Conformer Encoder and a cross-frame Conformer Encoder, each of which has 3 Conformer layers and 4 heads. The decision to reduce the Encoder size stems from the substantial GPU memory demands associated with multi-channel data processing. While this reduction may lead to a slight performance decline, it is offset by the advantages conferred by analyzing multi-channel data.

\subsection{Training and Inferring Details}
\subsubsection{Front-end pretraining}
Both the ResNet and the Conformer front-end models are pretrained on the VoxCeleb 2 dataset \cite{VoxCeleb}. For the ResNet model, we follow the training protocol mentioned in \cite{wang2020dku} and achieve 0.814\% EER on the Vox-O trial. For the Conformer model, we follow the training protocol mentioned in \cite{cai2023leveraging}, which has 0.65\% EER on the Vox-O trial.

\subsubsection{Training process}
At the begining, the front-end module is initialized from a pretrained ResNet34- or Conformer-based speaker embedding model, and other parts are randomly initialized. The training process contains three different stage:
\begin{itemize}
    \item Stage 1: Keep the front-end module frozen, and we only train the back-end module with the simulated data for 100,000 steps. The last model will be used for next stage.
    \item Stage 2: Unfreeze the front-end module and train the whole OTS-VAD model with 20\% of the real data and 80\% of the simulated data for 50,000 steps. The model will be validated every 500 steps and the best model with the lowest DER will be used for next stage.
    \item Stage 3: Continue to train the whole OTS-VAD model with only real training data for 50,000 steps. The model will be validated every 500 steps.
\end{itemize}
For DIHARD III, the real training data is DIHARD III development set. For Alimeeting, the real training data is Alimeeting training set. 

During training, the $L$ is set to 32s as mentioned in Section \ref{sec:method}, which means that the input data is a 32-second audio signal and will be split to two 16-second blocks, as shown in Fig. \ref{fig:TS-VAD-ResNet}. The model is optimized by Adam optimizer with a max learning rate of $10^{-4}$ at stage 1, $10^{-5}$ at stage 2, and $5\times 10^{-6}$ at stage 3. The learning rate was increased from zero linearly over the first 2000 updates, and then it was annealed to 0 using a cosine schedule. The acoustic features are 80-dimensional log Mel-filterbank energies with a frame length of 25ms and a frame shift of 10 ms. Therefore, the time resolution of acoustic features is $0.01s$. Then the features are downsampled by the front-end, and the time resolution of the frame-level embeddings is $0.08s$ for ResNet front-end and $0.02s$ for Conformer front-end. For Conformer front-end, we then employ two 1d convolution layers to reduce the time resolution by 4 times. Since Conformer and LSTM in the back-end do not change the sample rate of input, the output time resolution is also $0.08s$ for both the ResNet and Conformer front-end modules, respectively. 

\subsubsection{Inference process}
For inference, we tried different block length $l\in\{2s, 4s, 8s, 16s\}$ and block shift $m\in\{0.4s, 0.8s, 1.6s\}$. Note that the block shift is also the latency of the OTS-VAD model. The upper and lower thresholds are tuned by grid search with a fixed block length and shift, e.g., 16s and 0.8s, respectively. After finding the upper and lower thresholds with the lowest DER, we apply these two thresholds for inference with other different block length and shift. Finally, the output of each block is averaged and binarized with a threshold of 0.5 and converted to RTTM file for scoring. During inference process, we need an additional VAD module to remove the silence regions, and oracle VAD is employed for simplicity in our experiments. 

Both two different inference processes mentioned in Fig. \ref{fig:inference1} and \ref{fig:inference2} are evaluated. For the buffer-based inference process, we only evaluate it on the DIHARD III dataset in Fig. \ref{fig:inference1}. The reason is that the average duration of Alimeeting dataset is about 30 minutes, which requires large memory to saving the buffer and long time for inference. For the accumulation-based inference process, we provide comprehensive evaluation for both datasets.

\subsubsection{Evaluation metric}
The OTS-VAD model is evaluated on the evaluation set of both datasets, where we use the Diarization Error Rate (DER) and Jaccard Error Rate (JER) as the evaluation metric. For the DIHARD III dataset, we follow the evaluation protocol in the DIHARD III challenge \cite{ryant2020third}, where no forgiveness collar will be applied to the reference segments prior to scoring. For Alimeeting, we follow the evaluation protocol in the Alimeeting challenge \cite{M2MeT}, where a forgiveness collar of 0.25 is employed. All overlapping speech will be evaluated. 

\begin{table*}[htbp!]
  \caption{ ResNet front-end DER (\%) and JER (\%) on different datasets with the accumulation-based inference process in Fig. \ref{fig:inference2}. }
  \label{tab:results_resnet_infer2}
  \centering
  \begin{tabular}[c]{lccccccccccc}
    \toprule
     \multirow{4}*{\textbf{Dataset}} & \multirow{4}*{\textbf{Pretraining}} & \multirow{4}*{\textbf{Front-end}} & \multirow{4}*{\textbf{Block shift}} & \multicolumn{8}{c}{\textbf{Block length}}  \\
     \cmidrule(lr){5-12} 
     & & & & \multicolumn{2}{c}{2s} & \multicolumn{2}{c}{4s} & \multicolumn{2}{c}{8s} & \multicolumn{2}{c}{16s} \\
     \cmidrule(lr){5-6} \cmidrule(lr){7-8}  \cmidrule(lr){9-10} \cmidrule(lr){11-12}  
     & & & & DER & JER & DER & JER & DER & JER & DER & JER \\
     
   \midrule
	\multirow{7}*{{DIHARD III Eval}} & \multirow{7}*{\cmark} & \multirow{3}*{ResNet}
	&   0.4s & 15.80 & 33.95 & 14.53 & 32.46 & \textbf{13.65} & 33.00 & 13.65 & 33.22 \\
	&&& 0.8s & 16.00 & 34.16 & 14.51 & 33.13 & 13.68 & 32.92 & 14.07 & 33.14 \\
	&&& 1.6s & 17.44 & 37.74 & 15.17 & 35.52 & 14.39 & 35.09 & 14.36 & 34.84 \\
 \cmidrule(lr){3-12} && \multirow{3}*{Conformer}
    &   0.4s & 19.34 & 36.27 & 17.18 & 35.65 & 17.90 & 37.46 & 17.63 & 38.84 \\
	&&& 0.8s & 19.00 & 36.03 & \textbf{16.33} & 34.78 & 16.56 & 36.71 & 17.24 & 38.56 \\
	&&& 1.6s & 18.58 & 37.55 & 16.89 & 36.98 & 16.88 & 39.11 & 17.23 & 40.51 \\
   \midrule
	\multirow{7}*{{DIHARD III Eval}} & \multirow{7}*{\xmark} & \multirow{3}*{ResNet}
	&   0.4s & 22.70 & 43.73 & 18.10 & 39.26 & \textbf{15.80} & {36.67} & 16.79 & 39.47 \\
	&&& 0.8s & 21.52 & 43.23 & 17.26 & 38.10 & 16.14 & 37.27 & 16.46 & 38.38 \\
	&&& 1.6s & 22.27 & 44.65 & 17.56 & 38.90 & 16.39 & 37.87 & 16.15 & 38.54 \\
 \cmidrule(lr){3-12} && \multirow{3}*{Conformer}
    &   0.4s & 30.21 & 47.96 & 22.43 & 43.48 & 20.95 & 44.07 & 21.65 & 47.07 \\
	&&& 0.8s & 27.51 & 47.39 & 21.91 & 42.77 & \textbf{20.05} & 43.84 & 21.10 & 47.24 \\
	&&& 1.6s & 27.35 & 50.09 & 21.61 & 43.84 & 20.18 & 44.75 & 21.76 & 47.94 \\
   \midrule
	\multirow{7}*{{Alimeeting Eval (1st Channel)}} & \multirow{7}*{\cmark} & \multirow{3}*{ResNet}
	&   0.4s & 16.77 & 25.66 & 10.22 & 21.92 & 7.26 & 18.00 & \textbf{5.37} & 15.32 \\
	&&& 0.8s & 15.92 & 26.83 & 10.27 & 21.93 & 8.57 & 19.85 & 5.67 & 15.80 \\
	&&& 1.6s & 12.60 & 23.41 & 10.06 & 22.26 & 8.62 & 20.06 & 8.43 & 19.75 \\
    \cmidrule(lr){3-12} && \multirow{3}*{Conformer}
    &   0.4s & 24.74 & 36.59 & 13.60 & 23.17 & 7.90 & 18.08 & \textbf{5.05} & 13.91 \\
	&&& 0.8s & 27.92 & 39.47 & 16.72 & 30.03 & 9.95 & 22.05 & 6.61 & 16.81 \\
	&&& 1.6s & 32.06 & 45.03 & 17.99 & 32.60 & 9.58 & 21.88 & 8.78 & 21.20 \\
  \midrule
	\multirow{7}*{{Alimeeting Test (1st Channel)}} & \multirow{7}*{\cmark} & \multirow{3}*{ResNet}
	&   0.4s & 15.62 & 26.40 & 11.46 & 21.14 & 12.39 & 19.05 & 6.87 & 14.65 \\
	&&& 0.8s & 13.84 & 25.26 & 13.74 & 21.26 &  8.86 & 14.82 & \textbf{4.96} & 12.54 \\
	&&& 1.6s & 14.32 & 27.77 & 11.25 & 20.50 &  7.69 & 13.81 & 5.46 & 13.57 \\
 \cmidrule(lr){3-12} && \multirow{3}*{Conformer}
    &   0.4s & 28.79 & 37.83 & 20.20 & 26.13 & 10.93 & 18.22 & \textbf{6.38} & 13.89 \\
	&&& 0.8s & 31.21 & 39.74 & 17.62 & 24.21 & 10.59 & 18.02 & 7.29 & 15.42 \\
	&&& 1.6s & 39.02 & 47.57 & 17.37 & 28.13 & 12.41 & 21.97 & 9.66 & 19.31 \\
   \midrule
	\multirow{3}*{{Alimeeting Eval (Multi Channels)}} & \multirow{3}*{\cmark} & \multirow{3}*{ResNet}
	&   0.4s & 16.68 & 28.90 & 9.32  & 19.31 & 8.97 & 19.09 & 5.26 & 13.05 \\
	&&& 0.8s & 16.27 & 27.67 & 8.77  & 18.03 & 9.06 & 19.79 & \textbf{4.71} & {12.03} \\
	&&& 1.6s & 12.38 & 23.81 & 10.51 & 22.51 & 9.22 & 20.11 & 4.81 & 12.17 \\

   \midrule
	\multirow{3}*{{Alimeeting Test (Multi Channels)}} & \multirow{3}*{\cmark} & \multirow{3}*{ResNet}
	&   0.4s &  8.69 & 19.10 &  6.48 & 15.65 & 5.50 & 13.80 & 4.97 & {12.16} \\
	&&& 0.8s & 10.70 & 21.25 &  9.38 & 18.15 & 8.08 & 13.05 & \textbf{4.50} & 12.95 \\
	&&& 1.6s & 12.61 & 23.31 & 11.14 & 19.80 & 7.07 & 13.04 & 4.50 & 12.60 \\
     \bottomrule
     \end{tabular}
\end{table*}
\section{Experimental Results and Discussion}

\subsection{Detailed performance with different block length and shift}
Tab. \ref{tab:results_resnet_infer1} and \ref{tab:results_resnet_infer2} shows the results of the OTS-VAD model with different front-end on DIHARD III dataset and Alimeeting dataset with different inference strategy as mentioned in Fig. \ref{fig:inference1} and Fig. \ref{fig:inference2}, where we evaluated the model with different block length and shift. 

Overall, with the buffer-based inference process, the best DER performance on DIHARD III dataset is 13.31\%. For the first channel of alimeeting dataset, the lowest DER is 5.05\% and 4.96\% for Eval and Test set, respectively. And, if incorporated with the mutil-channel data, the DER will be further reduced to 4.71\% and 4.50\% for Alimeeting Eval and Test set, respectively.

For different front-end modules, as Tab. \ref{tab:results_resnet_infer1} and \ref{tab:results_resnet_infer2} show, ResNet front-end usually achieves better DER and JER scores compared to the Conformer front-end across most block lengths and block shifts. ResNet front-end achieves its best DER score at a block length of 8s or 16s.

For different inference processes, the buffer-based inference process shows a better performance than the accumulation-based inference process does in most conditions. The choice of the inference process might be application-specific, depending on the specific goals and constraints of the task. For the DIHARD III dataset, it only shows a tiny performance degradation, e.g., 13.65\% compared with 13.31\%.

Block length and shift also matter for a better performance. For DIHARD III, the best performance in Tab. \ref{tab:results_resnet_infer1} and \ref{tab:results_resnet_infer2} are typically achieved with an 8s or 16s block length. Conversely, for Alimeeting, longer block lengths consistently yield improved and stable performance. This contrast may be attributed to the heterogeneous domains of the DIHARD III dataset, making hyper-parameter tuning more difficult and resulting in less stable outcomes. Furthermore, inference with lower block shifts exhibit enhanced stability, while longer block shifts introduce greater performance variability.

We also evaluate the multi-channel data on the cross-channel-based model, which improves the performance on the Eval and Test set by 14\% and 9\%, repectively. In addition, it can provide a more stable performance under different condition, e.g., it can still show a satisfying performance with block length of 2s on the Test set. Actually, compare with the offline TS-VAD \cite{wang2022cross}, the improvement of multi-channel extension in online VAD is moderate, the reason is that we reduce the Encoder size to ensure enough GPU memory for training (3 layers 4 heads v.s. 6 layers 8 heads). We believe the performance will be much better if the same Encoder is employed for multi-channel extension. 

% Tab. \ref{tab:results_resnet_infer1} shows the OTS-VAD results with the embedding buffer-based inference process mentioned in Fig. \ref{fig:inference1}. Given the block length of 16s and block shift of 0.8s, the ResNet-based OTS-VAD achieves the best performance on the DIHARD III evaluation dataset, which has a DER of 13.31\% and a JER of 33.83\%. 
% For the single-channel data of the Alimeeting dataset, it achieves a DER of 5.35\% on Eval set and a DER of 12.34\% on the Test set. However, as it takes very long time for inference of the multi-channel data, we do not present the results for the multi-channel data of the Alimeeting dataset. 

% Tab. \ref{tab:results_resnet_infer2} shows the OTS-VAD results with the total embedding-based inference process mentioned in Fig. \ref{fig:inference2}. 

\begin{figure*}[htbp!]
    \centering
  \subfloat[R8009\_M8018\_MS809\label{1a}]{%
       \includegraphics[width=0.24\linewidth]{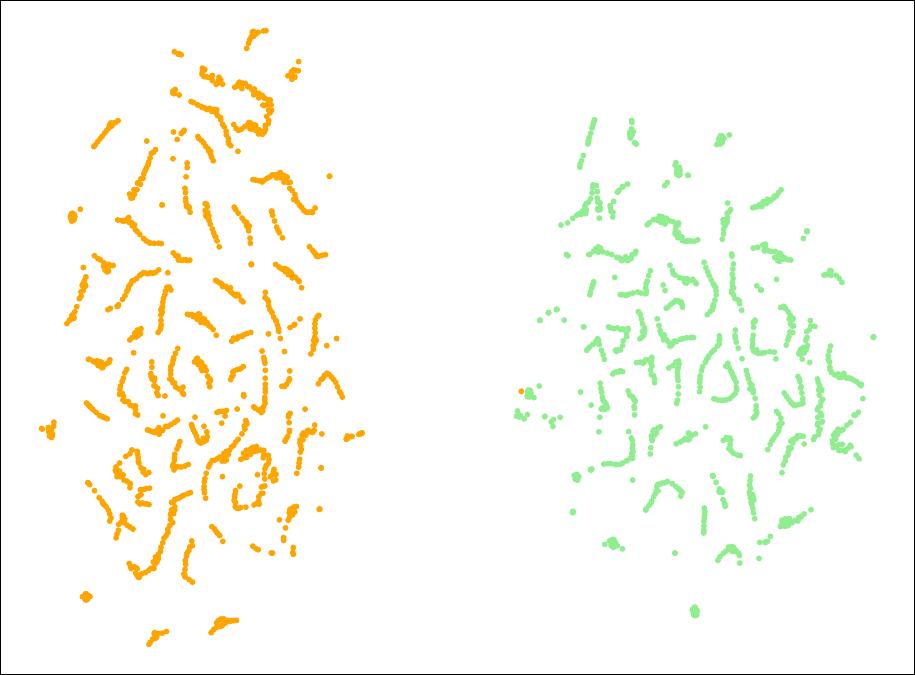}}
       \hfill
  \subfloat[R8001\_M8004\_MS801\label{1b}]{%
        \includegraphics[width=0.24\linewidth]{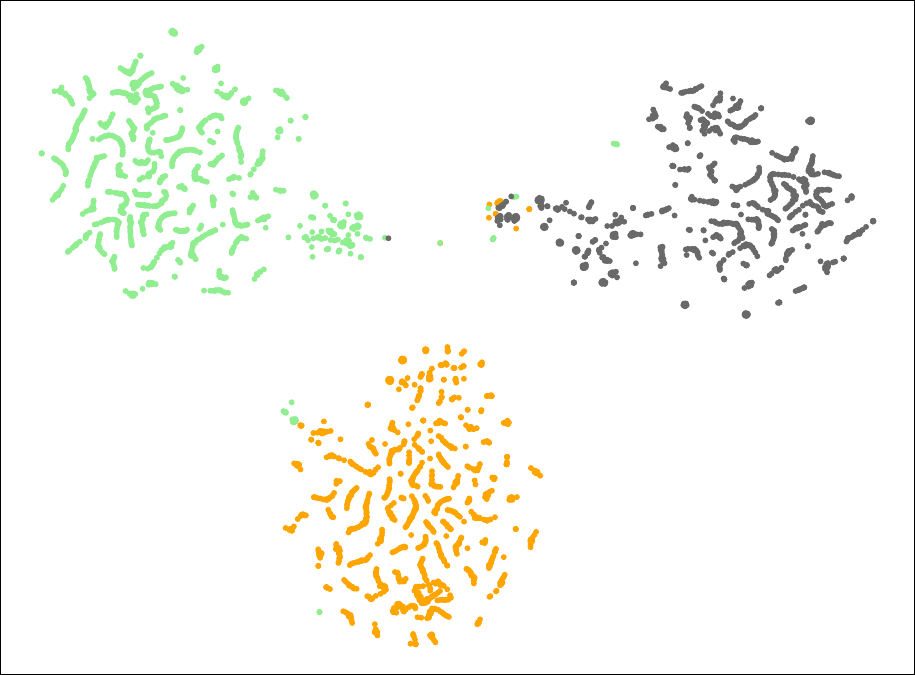}}
        \hfill
  \subfloat[R8008\_M8013\_MS807\label{1c}]{%
        \includegraphics[width=0.24\linewidth]{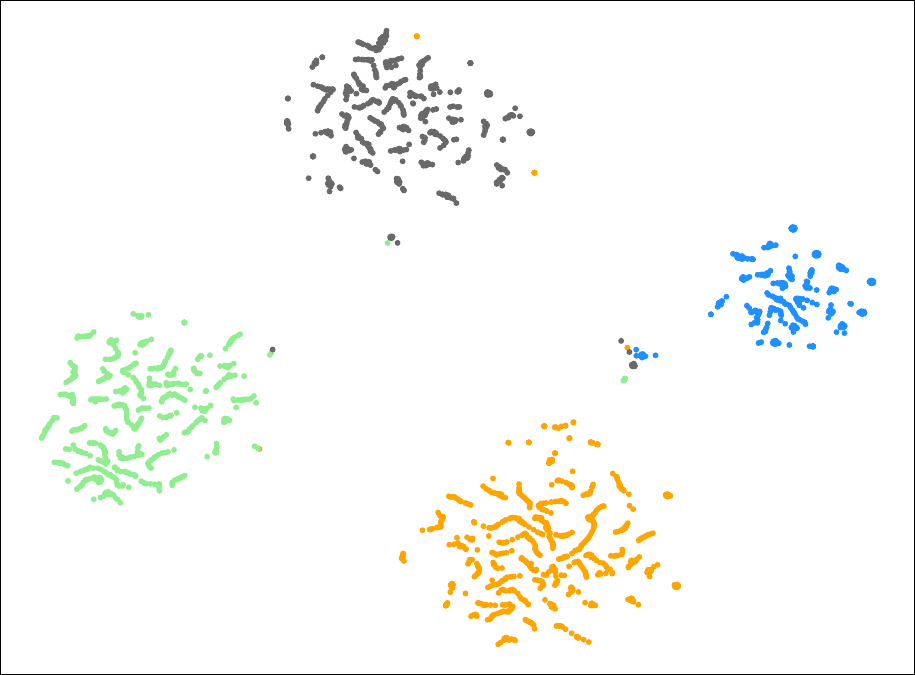}}
        \hfill
  \subfloat[DH\_EVAL\_0042\label{1d}]{%
        \includegraphics[width=0.24\linewidth]{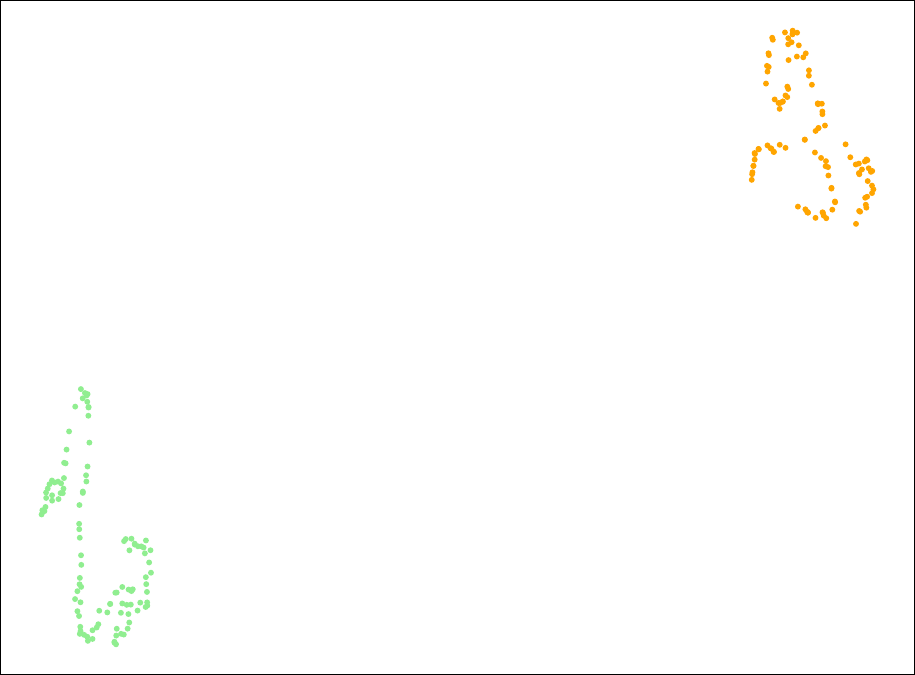}}
    \\
    \subfloat[DH\_EVAL\_0183\label{2a}]{%
       \includegraphics[width=0.24\linewidth]{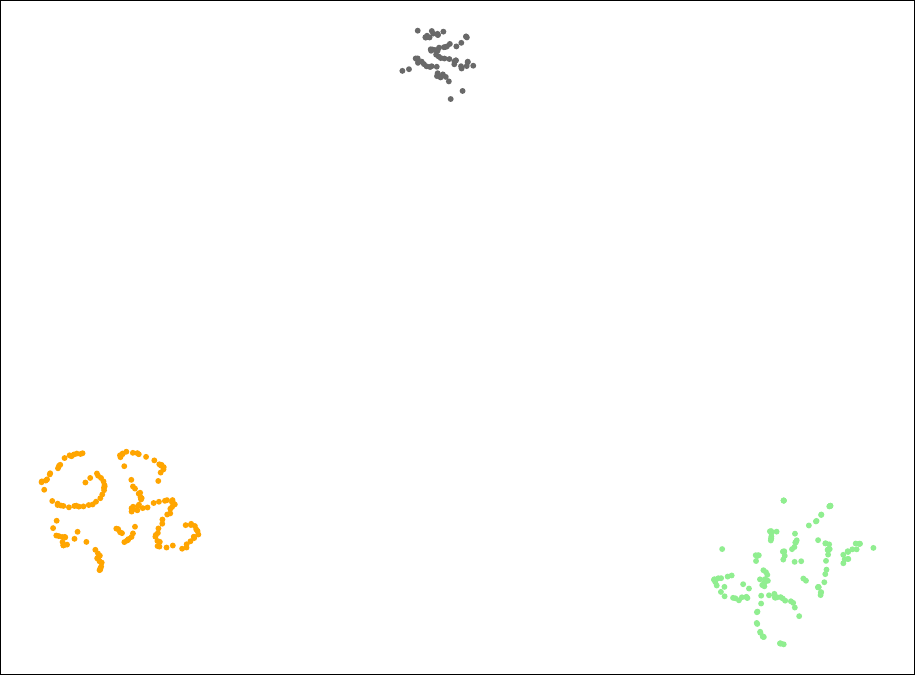}}
       \hfill
  \subfloat[DH\_EVAL\_0039\label{2b}]{%
        \includegraphics[width=0.24\linewidth]{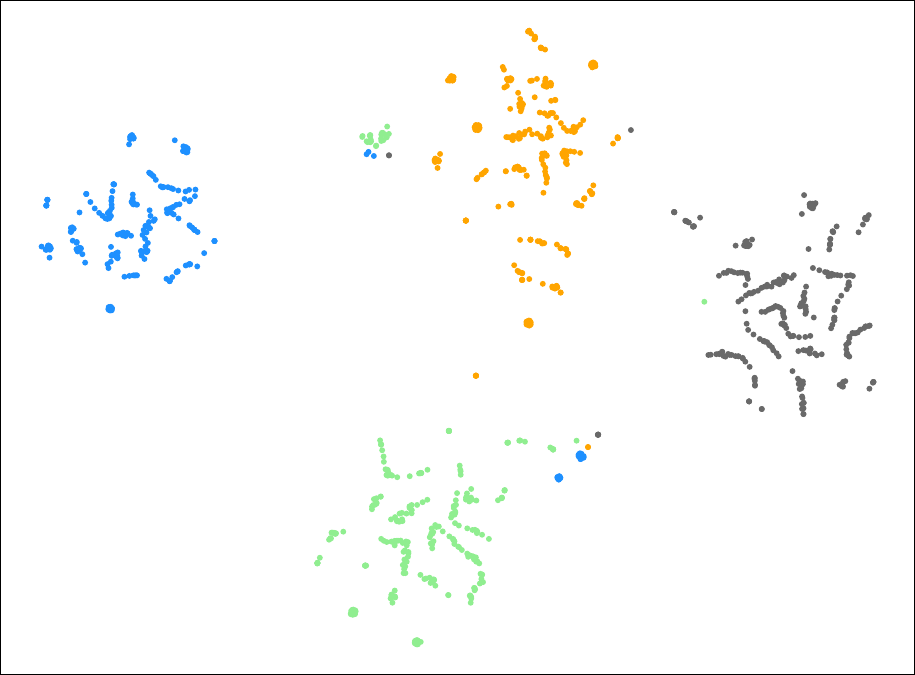}}
        \hfill
  \subfloat[DH\_EVAL\_0044\label{2c}]{%
        \includegraphics[width=0.24\linewidth]{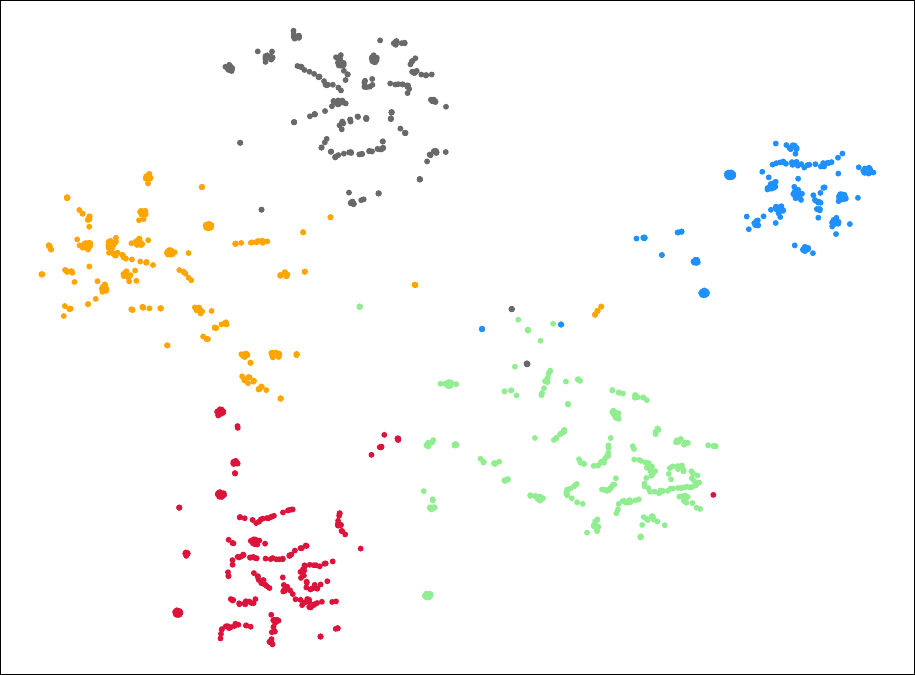}}
        \hfill
  \subfloat[DH\_EVAL\_0196\label{2d}]{%
        \includegraphics[width=0.24\linewidth]{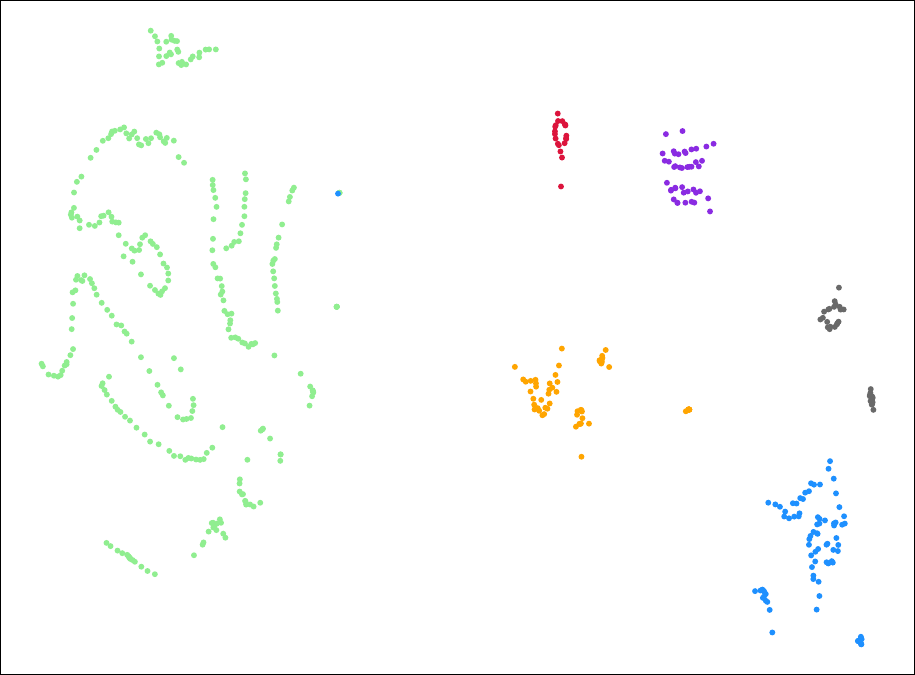}}
  \caption{t-SNE visualization of the target speaker embedding in each 16-second block during inference process, where the cosine is employed as the distance metric. (a), (b) and (c) are the samples from Alimeeting Eval dataset, each of which contains two, three and four speakers, respectively. (d), (e), (f), (g) and (h) are the samples from DIHARD III dataset, each of which contains $2 \sim 6$ speakers, respectively.}
  \label{fig:SpeakerEmbedding} 
\end{figure*}

\subsection{Comparison with other offline and online systems}

Tab. \ref{tab:comparison_dihard} shows the comparative analysis of various speaker diarization systems' performances on the DIHARD III dataset using the metric of Diarization Error Rates (DERs), where both offline or online systems with oracle VAD are included. For offline system, the Seq2seq TS-VAD is the SOTA system which achieves a DER of 10.77\%. For online systems, ResNet-based OTS-VAD shows the best performance with a DER of 13.31\%.

Tab. \ref{tab:comparison_alimeeting} shows the comparison with other systems on the Alimeeting datset. For offline system, we show the official baseline \cite{M2MeT} and the winner's system \cite{wang2022cross}. For online system, we do not find other online system evaluated on the Alimeeting dataset, but we can directly compare it with the offline TS-VAD system. For the Alimeeting Eval set, both the single-channel and multi-channel OTS-VAD show a similar performance with the single-channel TS-VAD. For Alimeeting Test set, the single-channel TS-VAD seems to be overfitted, but the multi-channel OTS-VAD still achieves a good performance compared with the offline multi-channel TS-VAD system.

\newcommand*{\MyIndent}{\hspace*{0.3cm}}%
\begin{table}[htbp!]
  \caption{ DERs (\%) of different systems on DIHARD III dataset, where oracle VAD is employed. }
  \label{tab:comparison_dihard}
  \centering
  \begin{tabular}[c]{lcccc}
    \toprule
     & &\multicolumn{2}{c}{\textbf{\# speakers}} & \\
     \cmidrule(lr){3-4} 
    Methods & Latency & $\le 4$ & $\ge 5$ & Total \\     
    \midrule
    \textbf{Offline} \\
    \MyIndent VBx + resegmentation \cite{bredin21_interspeech} & - & N/A & N/A &16.2\\
    \MyIndent Seq2Seq TS-VAD \cite{cheng2023target} & - & 6.65 & 25.90 & 10.77\\
    \midrule
    \textbf{Online} \\
        \MyIndent Zhang et al. \cite{zhang22_odyssey} & 0.5s & N/A & N/A & 19.57 \\
        \MyIndent Core samples selection \cite{yue22b_interspeech} & 2s & N/A & N/A & 19.3 \\
        \MyIndent EEND-GLA-Large \cite{OnlineEENDGLA} & 1s & 8.85 & 38.86 & 14.70 \\
        \MyIndent ResNet-based OTS-VAD & 0.8s & \textbf{8.78} & \textbf{32.03} & \textbf{13.31} \\
    \bottomrule
    \end{tabular}
\end{table}

\begin{table*}[htbp!]
  \caption{ DERs (\%) of different systems on Alimeeting dataset, where oracle VAD is employed. \\ SC refers to single channel and MC refers to multi channel}
  \label{tab:comparison_alimeeting}
  \centering
  \begin{tabular}[c]{lccccccccc}
    \toprule
     & &\multicolumn{4}{c}{\textbf{Alimeeting Eval set}} &\multicolumn{4}{c}{\textbf{Alimeeting Test set}} \\
     \cmidrule(lr){3-6}\cmidrule(lr){7-10} Methods & Latency & 2 spk & 3 spk & 4 spk & All & 2 spk & 3 spk & 4 spk & All   \\     
    \midrule
    \textbf{Offline} \\
    \MyIndent Official baseline \cite{M2MeT} & - & 4.84 & 10.90 & 23.18 & 15.24 & 3.00 & 6.36 & 30.59 & 15.60\\
    \MyIndent SC TS-VAD \cite{wang2022cross} & - & 0.89 & 6.63 & 5.47 & 4.11 & - & - & - & - \\
    \MyIndent MC TS-VAD \cite{wang2022cross} & - & 0.61 & 3.16 & 3.05 & \textbf{2.25} & 0.15 & 5.19 & 4.37 & \textbf{2.98} \\
    \midrule
    \textbf{Online} \\
    \MyIndent ResNet-based SC OTS-VAD & 0.8s & 0.85 & 3.98 & 9.27 & 5.67 & 0.33 & 6.18 & 8.39 & 4.96 \\
    \MyIndent ResNet-based MC OTS-VAD & 0.8s & 0.77 & 4.04 & 7.45 & \textbf{4.71} & 0.43 & 11.60 & 4.73 & \textbf{4.50} \\
    \bottomrule
    \end{tabular}
\end{table*}

% \subsection{Influence of block length and shift}
% Compared with the results in Tab. \ref{tab:results_resnet_infer1} and \ref{tab:results_resnet_infer2}, we can find that inferring with longer block can usually show better performance, but there are not many differences between results of different block shifts. 

% For DIHARD III dataset, the OTS-VAD model can still work well given only 4-second block length, which does not show much performance degradation. In addition, model with 2-second block length can also provide satisfying results, which proves that the front-end ResNet can extract robust speaker representations with very short blocks. 

% For Alimeeting dataset, the performance of short blocks degrades rapidly. The reason may be that the average duration of the recordings in Alimetting dataset is about 30 minutes. Audio recording with longer duration may accumulate more errors with the relatively short blocks during inference process.

\subsection{Importance of pretraining}
We also evaluate if the speaker verification pretraining of the front-end module is necessary. As Tab. \ref{tab:results_resnet_infer1} and \ref{tab:results_resnet_infer2} show, the model trained from scratch has worse performance, especially for very short block length. This means that the front-end without speaker verification pretraining cannot generalize well for the frame-level embeddings. 

%In addition, Fig \ref{} shows the loss during training within stage 1, 2 and 3 on DIHARD III daraset, where epoch 1 to 100 is stage 1, epoch 101 to 200 is stage 2 and epoch 201 to 300 is stage 3. 

\subsection{Visualization of target speaker embedding}
To evaluate how the model learns the frame-level embeddings to form the target-speaker embedding, we also show the t-SNE of the ResNet-based target speaker embedding in each of the blocks with different number of speakers in Fig. \ref{fig:SpeakerEmbedding}, where Fig. \ref{1a}, \ref{1b} and \ref{1c} are from Alimeeting dataset with 2$\sim$4 speakers, and Fig. \ref{1d}$\sim$\ref{2d} are from DIHARD III dataset with 2$\sim$6 speakers. Each point in the figure are the mean embedding of the frame-level embedding in the blocks during inference stage. As Fig. \ref{fig:SpeakerEmbedding} shows, the target speaker embedding can be well separated even we do not specify an objective to training the embedding. 

\subsection{Training/inference time and real-time factor}
Our experiments are conducted on NVIDIA Geforce RTX 3090 GPU. The training time is about 2 days with two GPUs for each single-channel dataset and 4 days with 8 GPUs for multi-channel dataset. For the accumulation-based inference process, we use one NVIDIA Geforce RTX 3090 GPU. For the single-channel ResNet-based OTS-VAD, the inference time for a 16s block is about 0.064s. For the multi-channel ResNet-based OTS-VAD, the inference time for a 16s block is about 0.186s. Given that latency is the block shift, the real-time factors (RTFs) can be easily calculated as $\frac{t}{m}$, where $t$ is the average inference time for a block and $m$ is the block shift.
The RTF for single channel and multi-channel OTS-VAD with 0.4s block shift are 0.16 and 0.465 , respectively. This demonstrates that the proposed method is possible for online inference. 

\section{Conclusion}

In this paper, we introduce an innovative end-to-end framework, OTS-VAD, designed for online speaker diarization using a target speaker profile as a guide. Adapted from the offline TS-VAD framework, this online version utilizes self-generated target speaker embeddings for detecting speaker activity. Evaluated on multiple datasets, including DIHARD III and Alimeeting, OTS-VAD consistently exhibited strong performance across varied scenarios, attesting to its reliability. In addition, we expanded the model's capability to handle multi-channel data, enhancing its resilience to different block lengths and shifts. We also transitioned from the ResNet front-end to the more efficient Conformer Encoder, which reduces the model size with little performance degradation. Our results show that OTS-VAD not only meets but in some cases surpasses state-of-the-art benchmarks, drawing close to certain offline speaker diarization systems. Its real-time factor suggests potential suitability for online applications. Given its ability to self-generate target speaker embeddings and produce the diarization results, OTS-VAD can be considered as a fully end-to-end speaker diarization system.

However, the model also has its limitations. First, it currently lacks an integrated voice activity detection (VAD) component, which could limit its applicability to real-world scenarios. Additionally, its relatively large size and extended CPU inference time present challenges, though techniques like quantization could mitigate these issues. Looking ahead, our future work will focus on incorporating a VAD module and refining the model, both in size and inference strategy, to better suit real-world data applications.

% or


%\appendices

%\section{}

%\section*{Acknowledgment}

%The authors would like to thank... \cite{IEEEexample:book_typical}

\ifCLASSOPTIONcaptionsoff
  \newpage
\fi



\bibliographystyle{IEEEtran}
{\footnotesize
\bibliography{mybib}}
\end{document}